\documentstyle[prl,twocolumn,psfig,floats,aps]{revtex}
\begin{document}
\wideabs{
\draft
\title{Destruction of localized electron pairs above the
magnetic-field-driven superconductor-insulator transition in
amorphous InO films}
\author{V.F.~Gantmakher\thanks{e-mail: gantm@issp.ac.ru},
M.V.~Golubkov, V.T.~Dolgopolov, G.E.~Tsydynzhapov, and A.A.~Shashkin}
\address{Institute of Solid State Physics, Russian Academy of
Sciences, 142432 Chernogolovka, Russia}
\maketitle

\begin{abstract}
We have investigated the field-induced superconductivity-destroying
quantum transition in amorphous indium oxide films at low
temperatures down to 30~mK. It has been found that, on the high-field
side of the transition, the magnetoresistance reaches a maximum and
the phase can be insulating as well as metallic. With further
increasing magnetic field the film resistance drops and approaches in
the high-field limit the resistance value at transition point so that
at high fields the metallic phase occurs for both cases. We give a
qualitative account of this behavior in terms of field-induced
destruction of localized electron pairs.
\end{abstract}
}

The theoretical description of the zero-field and field-induced
quantum superconductor-insulator transitions (SIT) in a 2D
superconductor is based on a concept of electron pairs which are
delocalized on the superconducting side and localized on the
insulating side of transition \cite{Grins,Fisher,Girvin}. According
to Refs.~\cite{Grins,Fisher,Girvin}, the temperature dependence of
the film resistance near the field-induced SIT is controlled by
deviation $\delta=B-B_c$ from the critical field $B_c$ and the most
specific among perceptible features of SIT is fan-like set of
resistance-vs-temperature curves $R_\delta (T)$. Such a set is
expected to collapse onto a single curve as a function of scaling
variable $\delta/T^{1/y}$, where $y$ is the critical index, see
review \cite{QPT}. Many of the SIT studies were performed on
amorphous In$_2$O$_x$ ($x<3$) films whose conductivity is caused by
oxygen deficiency compared to fully stoichiometric insulating
compound In$_2$O$_3$: by changing the oxygen content one can cover
the range from a superconductor to an insulator and thus realize the
zero-field SIT. On the insulating side of this SIT, observation was
reported of the activation behavior of the resistance
$R\propto\exp(T_0/T)^p$ with $p=1$ (Arrhenius law) and activation
energy $T_0$ tending to zero as the phase boundary is approached
\cite{Ovadyahu}. It was found later that switching a magnetic field
results in decreasing the resistance and weakening its temperature
dependence from the Arrhenius law to the Mott law with exponent
$p=1/4$ \cite{Golubkov}. This was explained in Ref.~\cite{Golubkov}
by magnetic-field-caused suppression of the binding energy $\Delta$
of localized electron pairs which manifests as a gap at the Fermi
level.

Field-induced SIT is realized on the superconducting side of
zero-field SIT. It is indicated by fan-like structure of experimental
curves $R_\delta (T)$ such that, in accordance with the scaling
analysis, the expected collapse is indeed the case \cite{Hebard}.
Above the field-induced SIT, the existence of two insulating phases
was postulated based on results of Hall measurements \cite{Ruel};
however, temperature dependences of the resistance of these phases
were not studied. Reversal of a zero-bias peak in the differential
resistance at the critical field $B_c$ was observed and attributed to
the granular structure of films \cite{Kim}.

Here, we investigate the phase on the high-field side of SIT where
the occurrence of localized electron pairs is predicted. We find
that, while this phase can be both insulating and metallic, in the
high-field limit the system always enters the metallic phase. This is
interpreted as field-caused breaking of localized electron pairs.

\begin{table}
\vbox{
\caption{Parameters of two states of the sample. $R_r$ is the
resistance at room temperature, the values of $R_c$ and $B_c$ are
determined by means of scaling analysis as described in
Ref.~\protect\cite{we}.}
\begin{tabular}{l|ccc}
State&$R_r$, k$\Omega$&$R_c$, k$\Omega$&$B_c$, T\\
\tableline
1&3.4&7.5&2\\
2&3.0&9.2&7.2\\
\end{tabular}
\label{table}
}
\end{table}
The experiments were performed on a 200~\AA\ thick amorphous
In$_2$O$_x$ ($x<3$) film without pronounced granularity as was
checked by the absence of quasireentrant transition, i.e., the
absence of minimum on dependences $R_\delta (T)$ at low temperatures
\cite{Gold}. The oxygen content $x$ could be reversibly altered by
heat treatment; all experimental procedures are described in detail
in Ref.~\cite{Golubkov}. Assuming for the sake of simplicity that the
film disorder remains unchanged during heat treatment, the quantity
$x$ controls the carrier density $n$ and, then, it is the variation
of $n$ that causes zero-field SIT. Two states of the film were
studied with parameters listed in Table~\ref{table}. Under the above
assumption, the carrier density in a state should be inversely
proportional to its room temperature resistance. Hence, state 2 is
farther from the zero-field SIT and deeper in the superconducting
phase as compared to state 1. The magnetoresistance of both states
was measured in Oxford TLM-400 dilution refrigerator in the
temperature range 1.2~K to 30~mK using a four-terminal lock-in
technique at a frequency of 10~Hz. The current across the sample was
equal to 5~nA and corresponded to the linear regime of response. The
measurement runs were made by sweeping magnetic field at fixed
temperature.

Our preceding study has confirmed the existence of
magnetic-field-tuned quantum SIT in such films and revealed that this
phenomenon is more general compared to the one considered in
Ref.~\cite{Fisher}. Particularly, to attain collapse of data
$R_\delta (T)$ in the vicinity of transition against scaling variable
$\delta/T^{1/y}$ account should be taken, e.g., of the temperature
dependence of the critical resistance $R_c$, which gives rise to the
appearance of a linear term in $T$ in dependences $R_\delta (T)$
\cite{we}.

\begin{figure}
\vbox{
\psfig{file=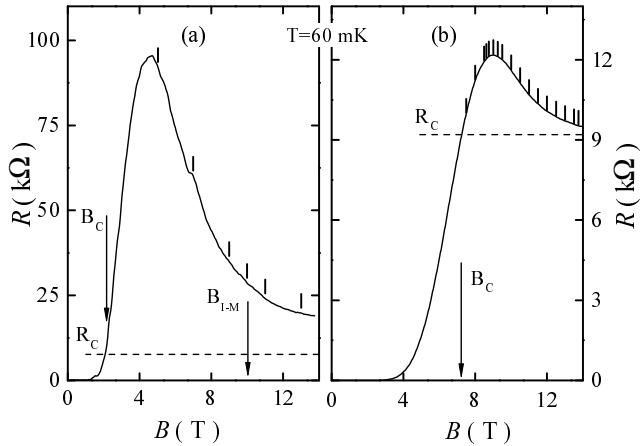,width=\columnwidth,clip=}
\caption{Magnetoresistance of the film in state 1 (a) and in state 2
(b). The critical $R_c$ and $B_c$ values at $T=0$ are indicated. Also
shown is the position of metal-insulator transition, $B_{I-M}$,
determined from Fig.~\protect\ref{RT}. The temperature dependences of
the resistance are analyzed at fields marked by vertical bars.}
\label{MR}
}
\end{figure}
Fig.~\ref{MR} displays the magnetoresistance traces for both states
of the film at a temperature of 60~mK. The critical field $B_c$ and
resistance $R_c$ at $T=0$ (Table~\ref{table}) are determined with the
help of scaling analysis as described in detail in Ref.~\cite{we}.
One can see from the figure that with increasing field the
magnetoresistance for both of the states reaches a maximum $R_{max}$
above $B_c$ and then drops so that in the high-field limit it
approaches the value of $R_c$. The relative value of maximum
$R_{max}/R_c$ is considerably larger for state 1 which is closer to
the zero-field SIT; moreover, the phase right above $B_c$ is
insulating in state 1 and metallic in state 2, as will be shown
below.

\begin{figure}
\vbox{
\psfig{file=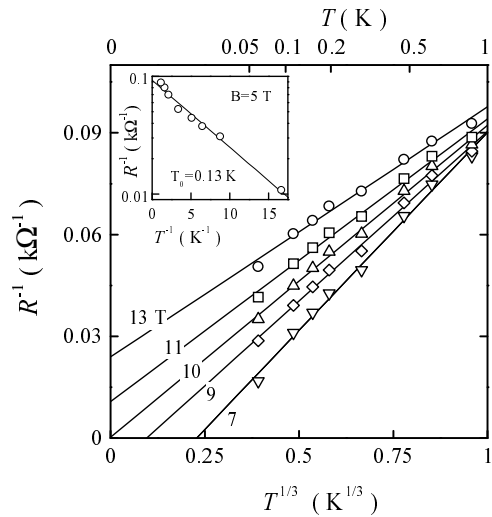,width=\columnwidth,clip=}
\caption{Temperature dependence of the high-field conductance of
state 1 at various magnetic fields. Arrhenius plot of the conductance
at $B=5$~T is displayed in the inset.}
\label{RT}
}
\end{figure}
The vertical bars in Fig.~\ref{MR} mark the magnetic field values at
which the temperature dependence of resistance is analyzed. The
results of such an analysis for state 1 are represented in
Fig.~\ref{RT}. At fields near the resistance maximum $R(T)$ follows
the activation behavior as expected for an insulator (inset to
Fig.~\ref{RT}). However, at higher fields the activation law does not
hold, nor do the logarithmic corrections normally observed in 2D
metals \cite{Bergmann}. That is why we examine the film resistance
over the field range 7 to 13~T in terms of 3D material behavior in
the vicinity of metal-insulator transition \cite{Imry,Zverev}:

\begin{equation}
\sigma(T)=a+bT^{1/3} \qquad b>0,\label{scale}\end{equation}
where the sign of the parameter $a$ discriminates between a metal and
an insulator at $T\rightarrow 0$. If $a>0$, it yields zero
temperature conductivity $\sigma(0)=a$, whereas the negative $a$
points to activated conductance at lower temperatures.

We emphasize that we judge about the transport properties at $T=0$ as
obtained by extrapolation from above 30~mK. Bearing this in mind, we
determine from Fig.~\ref{RT} the field $B_{I-M}\approx 10$~T of
metal-insulator transition for state 1. So, a conclusion of
Ref.~\cite{Ruel} about the presence of two phases above the SIT is
confirmed. Yet, in contrast to Ref.~\cite{Ruel}, we find that their
phase boundary is not near the resistance maximum but at appreciably
higher field and that the high-field phase is metallic
(Fig.~\ref{MR}).

\begin{figure}
\vbox{
\psfig{file=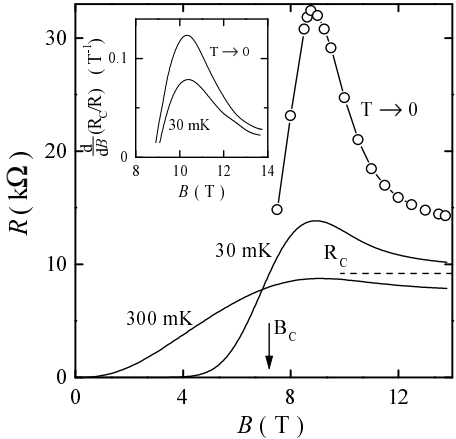,width=\columnwidth,clip=}
\caption{Magnetoresistance of the film in state 2 at $T=30$ and
300~mK, and for $T\rightarrow 0$ as obtained from extrapolations
(circles) in accordance with Eq.~(\protect\ref{scale}). The critical
field and resistance are indicated. The field derivatives of $R_c/R$
for $T\rightarrow 0$ and $T=30$~mK are compared in the inset.}
\label{RB0}
}
\end{figure}
For state 2 the parameter $a$ is positive over the entire field range
7.5 to 14~T above $B_c$ so that there is no insulating phase. The
corresponding field dependence of $R_0\equiv R_\delta (0)=1/a$ is
presented in Fig.~\ref{RB0} alongside with the experimental curves
$R(B)$ at 30 and 300~mK. Although the extrapolation is far, the
tendency for the lowest temperature data to approach $R_c$ in the
high-field limit seems valid for the extrapolated dependence as well.

The rise of the resistance near the field-driven quantum SIT is in
agreement with theoretical ideas about localized electron pairs:
above $B_c$ it reflects the decrease of the pair localization length
$\xi_{loc}$ with increasing $B$ \cite{Grins,Fisher,Girvin}. That the
resistance reaches a maximum with further increasing $B$ was not
discussed theoretically so far. Nevertheless, a qualitative account
of the observed resistance drop with field can be given in terms of
pair breaking caused by magnetic field \cite{Golubkov}. In this case
the behavior of the system of depaired electrons is naturally
determined by their density $n_d(B)$: at low $n_d$ depaired electrons
are certainly localized whereas at sufficiently high $n_d$ a
metal-insulator transition may be expected. It is the second
conduction channel that allows interpretation of the observed
nonmonotonic dependences of magnetoresistance.

\begin{figure}
\vbox{
\psfig{file=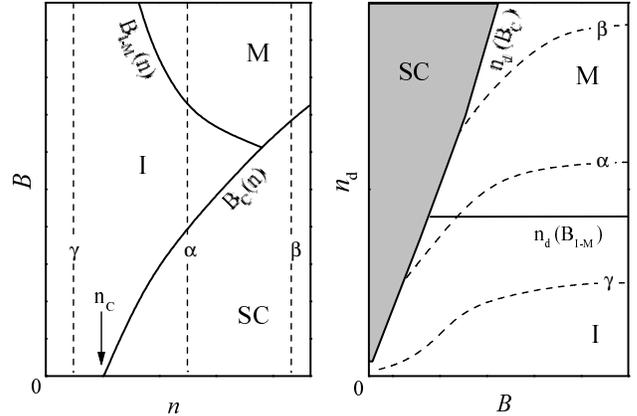,width=\columnwidth,clip=}
\caption{Schematic phase diagram of the observed transitions in the
($n,B$) and ($B,n_d$) planes. The evolution of states $\alpha$,
$\beta$, $\gamma$ with magnetic field is shown by dashed lines. In
shaded area the value $n_d$ is not defined.}
\label{DI}
}
\end{figure}
In agreement with experiment, Fig.~\ref{DI} schematically shows the
field behavior for three states of the sample. Two of these, $\alpha$
and $\beta$, that are selected in the superconducting phase above the
zero-field SIT at $n> n_c$ correspond to the studied states 1 and 2,
respectively. State $\gamma$ selected in the insulating phase at $n<
n_c$ corresponds to samples from Ref.~\cite{Golubkov}. With
increasing $B$ state $\alpha$ undergoes the field-induced SIT at
$B=B_c$ so that the depaired electrons available are localized at
that small density $n_d$ (Fig.~\ref{DI}). With further increase of
field the value of $n_d$ enhances due to localized pair breaking,
giving rise to the metal-insulator transition in the system of
depaired electrons. At $B\rightarrow\infty$, all electron pairs are
expected to be broken and so the value of $n_d$ should be equal to
the carrier density $n$. The different behavior of state $\beta$ is
due to larger density $n_d$ at the field-induced SIT because of
higher $n$ and $B_c$. As a result, the depaired electrons are already
delocalized at $B=B_c$ and thus the field range of the insulating
phase shrinks as the zero-field SIT is departed from, see
Fig.~\ref{DI}. Finally, state $\gamma$ approaches the metal-insulator
phase boundary with increasing $B$ but remains insulating for all
fields \cite{Golubkov}.

Thus, the concept of field-induced pair breaking requires the
additional assumption that a metal-insulator transition occurs in the
system of depaired electrons. Also, a theory
\cite{Grins,Fisher,Girvin} should be extended to include the
possibility of a direct superconductor-metal quantum transition.

Although the origin of localized electron pairs is still an open
question, a likely candidate for their breaking might be the
paramagnetic effect. In this case the pair breaking field should be
proportional to the binding energy of a pair $B^*=2\Delta/g\mu_B$,
where $g$ is the Land\'e factor. It is clear that the broad field
interval of the negative differential magnetoresistance points to
wide dispersion of the pair binding energies. To estimate the
distribution function $\nu(\Delta)$ for state 2 we presume for the
sake of simplicity that, at fields above the resistance maximum,
$R_0(B)$ at $T\rightarrow 0$ is inversely proportional to the density
of depaired electrons (the Drude limit)

\begin{equation}
R_c/R_0=n_d/n ,\label{RRc}\end{equation}
where $n$ is the carrier density in the metallic state at
$B\rightarrow\infty$, and $n_d$ is given by the formula

\begin{equation}
n_d=n-2\int\limits^\infty_{g\mu_B B/2}\nu(\Delta){\rm d}\Delta .
\label{n}\end{equation}
Then, it is easy to obtain the distribution function

\begin{equation}
\nu(\Delta)=\frac{2n}{g\mu_B}\frac{\rm d}{\rm d\it B}(R_c/R_0)
\Bigr|_{B=2\Delta/g\mu_B} .\label{nu}\end{equation}
The field derivative of the ratio $R_c/R_0$, which is proportional to
$\nu(B)$, is depicted in the inset of Fig.~\ref{RB0}. Its behavior is
similar to that of the field derivative of $R_c/R(T=30$~mK) in spite
of the far extrapolation to get $R_0(B)$, see Fig.~\ref{RB0}.

The fact that the distribution $\nu(\Delta)$ is broad allows us to
distinguish between two localization scenarios of electron pairs: (i)
the localization radius $\xi_{loc}$ is larger than the pair size
$\xi_0$; and (ii) $\xi_{loc} <\xi_0$. In the first case the binding
energy $\Delta$ is determined mainly by intrinsic factors and is
expected to be approximately the same for all pairs. In the opposite
case two electrons forming a pair are localized at separate sites and
so the binding energy of the pair depends crucially on local random
potential \cite{Golubkov}. This implies that the dispersion of
$\Delta$ is broad. Hence, the obtained data are likely to point to
the second scenario of localization. We note that the limit
$\xi_{loc} <\xi_0$ was assumed in a model of localized bipolarons
\cite{Alex}.

In summary, our study of the field-driven quantum SIT in amorphous
In$_2$O$_x$ films shows that, on the high-field side of the
transition, with increasing $B$ the film magnetoresistance reaches a
maximum and then drops approaching in the high-field limit the
resistance $R_c$ at transition point. We find that the high-field
phase is always metallic while the phase right above $B_c$ can be
both insulating and metallic, dependent on distance to the zero-field
SIT. The obtained experimental data can be understood within a model
of localized electron pairs if one includes (i) a concept of
field-caused pair breaking that presumes a metal-insulator transition
in the system of depaired electrons; and (ii) a meaning of
superconductor-metal quantum transition. That the negative
differential magnetoresistance is observed in a wide field region is
likely to point to broad dispersion of the binding energies of
localized electron pairs.

This work was supported by Grants RFBR~96-02-17497, RFBR~97-02-16829,
INTAS-RFBR~95-302 and by the Programme "Statistical Physics" from the
Russian Ministry of Sciences.


\begin{references}
\bibitem{Grins} M.P.A.~Fisher, G.~Grinshtein, and S.M.~Girvin, Phys.\
Rev.\ Lett.\ {\bf 64}, 587 (1990).
\bibitem{Fisher} M.P.A.~Fisher, Phys.\ Rev.\ Lett.\ {\bf 65}, 923
(1990).
\bibitem{Girvin} S.M.~Girvin, M.~Wallin, M.-C.~Cha, et al., Prog.\
Theor.\ Phys.\ Suppl.\ {\bf 107}, 135 (1992).
\bibitem{QPT} S.L.~Sondhi, S.M.~Girvin, J.P.~Carini, and D.~Shahar,
Rev.\ Mod.\ Phys.\ {\bf 69}, 315 (1997).
\bibitem{Ovadyahu} D.~Shahar, and Z.~Ovadyahu, Phys.\ Rev.\ B\ {\bf
46}, 10917 (1992).
\bibitem{Golubkov} V.F.~Gantmakher, M.V.~Golubkov, J.G.S.~Lok, and
A.K.~Geim, JETP {\bf 82}, 951 (1996).
\bibitem{Hebard} A.F.~Hebard, and M.A.~Paalanen, Phys.\ Rev.\ Lett.\
{\bf 65}, 927 (1990).
\bibitem{Ruel} M.A.~Paalanen, A.F.~Hebard, and R.R.~Ruel, Phys.\
Rev.\ Lett.\ {\bf 69}, 1604 (1992).
\bibitem{Kim} K.~Kim, and H.-L.~Lee, Phys.\ Rev.\ B\ {\bf 54}, 13152
(1996).
\bibitem{Gold} Y.~Liu, D.B.~Haviland, B.~Nease, and A.M.~Goldman,
Phys.\ Rev.\ B\ {\bf 47}, 5931 (1993).
\bibitem{we} V.F.~Gantmakher, M.V.~Golubkov, V.T.~Dolgopolov,
G.E.~Tsydynzhapov, and A.A.~Shashkin, cond-mat/9806244.
\bibitem{Bergmann} G.~Bergmann, Phys.\ Rep.\ {\bf 107}, 1 (1984).
\bibitem{Imry} Y.~Imry, and Z.~Ovadyahu, J.\ Phys.\ C\ {\bf 15}, L327
(1982).
\bibitem{Zverev} V.F.~Gantmakher, V.N.~Zverev, V.M.~Teplinskii, and
O.I.~Barkalov, JETP {\bf 76}, 714 (1993).
\bibitem{Alex} A.S.~Alexandrov, and N.F.~Mott, Rep.\ Prog.\ Phys.\
{\bf 57}, 1197 (1994).
\end{references}
\end{document}